      [1999/12/01 v1.4c Il Nuovo Cimento]
\documentclass{cimento}

\usepackage{graphicx}  
\title{Anomalously interacting extra neutral bosons}
\author{M.V.~Chizhov\from{ins:s}\from{ins:x}\ETC,
V.A.~Bednyakov\from{ins:x}
        \atque
J.A.~Budagov\from{ins:x}\thanks{In memory of Alexei Norairovich Sissakian
}}
\instlist{\inst{ins:s} CSRT, Faculty of Physics, Sofia University,
1164 Sofia, Bulgaria \inst{ins:x} DLNP, Joint Institute for Nuclear
Research, 141980, Dubna, Russia}
\PACSes{\PACSit{12.60.Cn,13.85.Fb,14.70.Pw}{}}
\begin{document}

\maketitle

\begin{abstract}
We study phenomenological consequences of the Standard Model
extension by the new spin-1 chiral fields with the internal quantum
numbers of the electroweak Higgs doublets.
      There are at least three different classes of theories,
      all motivated by the hierarchy problem, which predict
      new vector weak-doublets with masses not far from the electro-weak scale.
      We discuss resonance production of these neutral chiral $Z^*$ bosons
      at hadron colliders.
      The bosons can be observed as a Breit-Wigner resonance peak
      in the invariant dilepton mass distributions
      in the same way as the well-known 
      extra gauge $Z'$ bosons.
      This includes
      them into a list of
      very interesting objects for early searches with the first LHC data.
      Moreover, the $Z^*$ bosons
      have unique signatures in transverse momentum,
       angular and pseudorapidity distributions of the
      final leptons,
      which allow 
      to distinguish
     them from the other heavy neutral resonances.
\end{abstract}

\section{Introduction}

The method of the covariant derivatives leads to the unique minimal
form of the gauge bosons couplings to the fermions. Although the
gauge symmetry allows
anomalous interactions in the initial Lagrangian, all known
fundamental spin-1 bosons, photon, $W^\pm$, $Z$ and gluons, possess
only renormalizable minimal interactions with the known fermions. 
The anomalous interactions are
considered as effective ones. They are
generated on the level of the quantum loop corrections.
Usually they are proportional
to the additional square of a small coupling constant and can be
neglected in the first order approximation.

A different picture is realized at the low energy QCD domain, where
gluon and quark degrees of freedom are substituted by physical
hadronic states. The latter can be described by an effective field
theory. For example, spin-1 boson states, associated with the vector
fields, interact with baryons in all possible ways.
So, due to strong dynamics, the vector $\rho$ meson has both, comparable by the
magnitude, minimal and anomalous couplings with
$\bar{\psi}\gamma^\mu\psi$ and
$\partial_\nu(\bar{\psi}\sigma^{\mu\nu}\psi)$ currents,
respectively~\cite{rho}.
The both currents have the
same
quantum numbers $J^{PC}=1^{--}$ and since the parity and charge conjugation are
conserved in QCD they define the quantum numbers of the meson.

The axial-vector meson $a_1$ has different quantum numbers $1^{++}$,
which allow it also to have both
a minimal interaction with $\bar{\psi}\gamma^\mu\gamma^5\psi$
current and an anomalous interaction with
$\partial^\mu(\bar{\psi}\gamma^5\psi)$ current. 
But the most interesting
is another axial-vector meson $b_1$
with the quantum numbers $1^{+-}$.
Due 
to the latter the meson has only anomalous interaction with the
tensor current
$\partial_\nu(\bar{\psi}\sigma^{\mu\nu}\gamma^5\psi)$.
In fact, this QCD feature can be applied to the electroweak physics
as well. We will see how it plays a key role
below.

Let us assume that the electroweak gauge sector of the Standard
Model (SM) is extended by a doublet of new spin-1 {\em chiral\/}
bosons $\mbox{\boldmath$W$}^*_\mu$ with the internal quantum numbers
of the SM Higgs boson. They can originate, for example, from the
extensions of the SM such as Gauge-Higgs unification,
larger gauge groups~\cite{Gia} or technicolor models. However, due
to the lack of fully realistic models, the collider 
expectations for signals from these chiral bosons have not yet been
studied in details. Nevertheless, it is possible to point out
several model independent and unique signatures, which allow
to identify the production of such bosons at the hadron
colliders~\cite{proposal}.

Since the tensor current mixes the left-handed and the right-handed
fermions, which in the SM are assigned to the different
representations, the gauge doublet should have only anomalous
interactions:
\begin{equation}
    \label{master}
    {\cal L}^*=\frac{g}{M}\left(
                     \partial_\mu W^{*-}_\nu\;  \partial_\mu \overline{W}^{*0}_\nu
                 \right)\cdot
                 \overline{D_R}\;\sigma^{\mu\nu}\left(\hspace{-0.2cm}
                                                        \begin{array}{c}
                                                          U_L \\
                                                          D_L
                                                        \end{array}\hspace{-0.2cm}
                                                      \right)+
    \frac{g}{M}\left(
    \overline{U_L}\;  \overline{D_L}
                        \right)
                        \sigma^{\mu\nu}D_R\cdot
                        \left(\hspace{-0.2cm}
                          \begin{array}{c}
                            \partial_\mu W^{*+}_\nu \\
                            \partial_\mu W^{*0}_\nu \\
                          \end{array}\hspace{-0.2cm}\right),
\end{equation}
where $M$ is the boson mass, $g$ is the coupling constant of the
$SU(2)_W$ weak gauge group, and $U$ and $D$ generically denote
up-type and down-type leptons and
quarks~\footnote{Here we assume also universality of lepton and quark
couplings with different flavors.}. The bosons, coupled to the
tensor quark currents, are some types of {\em excited\/} states as
far as the only orbital angular momentum with $L=1$ contributes to
the total angular moment, while the total spin of the system is
zero. This property manifests itself in their derivative couplings
to fermions and in the different chiral structure of the interactions
in contrast to the minimal gauge interactions.

For simplicity in
(\ref{master})
we have introduced only interactions with the
down-type right-handed singlets, $D_R$. The coupling constant is
chosen in such a way that in the Born approximation all partial
fermionic decay widths of the well-known hypothetical $W'$ boson
with the SM-like interactions
\begin{equation}\label{W'}
    {\cal L}'_{CC}=\frac{g}{\sqrt{2}}\,W'^-_\mu\cdot\overline{D_L}\gamma^\mu
    U_L
    +\frac{g}{\sqrt{2}}\,\overline{U_L}\gamma^\mu D_L\cdot
    W'^+_\mu
\end{equation}
and the charged $W^{*\pm}$ boson with the same mass are identical.

In the same way as in many of the SM extensions several Higgs
doublets are  introduced the realistic model could include several
gauge doublets. Using the charge-conjugated doublet
\begin{equation}\label{W*c}
    \mbox{\boldmath $W$}^{*\,{\rm c}}_\mu=\left(
             \begin{array}{c}
               \overline{W}^{*0}_\mu \\
               -W^{*-}_\mu \\
             \end{array}
           \right)
\end{equation}
(or new ones with the hypercharges opposite to the $\mbox{\boldmath
$W$}^*_\mu$ doublet) it is possible to construct more complicated
models including up-type right-handed singlets, $U_R$, as well.

\section{The Model}
The minimal set of the chiral heavy bosons in the proposed extension
of the SM consists of the four spin-1 particles: the two charged
$W^{*\pm}$ states and the two neutral $CP$-even
$Z^*=(W^{*0}+\overline{W}^{*0})/\sqrt{2}$ and $CP$-odd
$\widetilde{Z}^*=(W^{*0}-\overline{W}^{*0})/\sqrt{2}\,i$
combinations. The corresponding Lagrangian for the neutral states
reads
\begin{equation}\label{Z*}
    {\cal
    L}^*_{NC}=\frac{g}{\sqrt{2}M}\left(\bar{D}\sigma^{\mu\nu}D\cdot\partial_\mu
    Z^*_\nu+i\bar{D}\sigma^{\mu\nu}\gamma^5 D\cdot\partial_\mu
    \widetilde{Z}^*_\nu\right).
\end{equation}
In the present paper we will discuss only the resonance production
of the neutral heavy bosons and their subsequent decay into a pair
of the light charged leptons. This process is the ``golden channel''
for
early discovery at the hadron colliders. However, in this case it
is impossible to discriminate the multiplicative quantum numbers of
the neutral bosons, namely $P$ and $C$, due to their identical
signatures. Therefore, in the following calculations we will
consider only one of them, for instance, $Z^*$ boson.

In order to compare the experimentally accessible distributions
between the tensor couplings and the vector ones, we introduce
topologically analogous but minimal gauge interactions of the $Z'$
boson
\begin{equation}\label{Z'}
    {\cal L}'_{NC}=\frac{g}{2}\,\bar{D}\gamma^\mu D\cdot Z'_\mu.
\end{equation}
In the Born approximation
eqs. (\ref{Z*}) and (\ref{Z'}) lead to the same cross-sections for
the hadro-production and decay of both neutral heavy bosons,
$Z^*(\widetilde{Z}^*)$ and $Z'$, when they have the same mass.
%
%
As we have assumed, the lepton and the quark couplings are
characterized by the same coupling constant, $g$, of the $SU(2)_W$
weak gauge group. So, the leptonic branching ratio is ${\cal
B}(Z^*\!/Z'\to\ell^+\ell^-)=1/12\approx 8\%$ and the total fermionic
decay width
\begin{equation}\label{Gf}
    \Gamma=\frac{g^2}{4\pi}\:M\approx 0.034\:M
\end{equation}
is around 3\% of the mass of the resonance.

All calculations we carried out
in the framework of the
CompHEP package~\cite{CompHEP}.
To this end a new model has been implemented, which includes
additional new bosons and their corresponding interactions.

\section{Numerical simulations for neutral bosons}

Up to now, the excess in the Drell--Yan process with high-energy
invariant mass of the lepton pairs remains the clearest indication
of a new heavy boson production at the hadron colliders on the early
stage. Therefore, we will concentrate on the production and decay of
the neutral bosons, where the full kinematics is experimentally
reconstructible. In the following we will use the CompHEP
package~\cite{CompHEP} for the numeric calculations of various
distributions for the inclusive processes
$pp\rightarrow\gamma/Z/Z^*\!/Z'\to\ell^+\ell^-$ with a CTEQ6L choice
for the proton parton distribution set. For both final leptons we
impose angular cuts relevant to the LHC detectors on the
pseudorapidity range $|\eta_\ell|<2.5$ and the transverse momentum
cuts $p_\sy{T}
> 20$~GeV.

\begin{figure}[htb]\center
\includegraphics[height=7cm,width=9.2cm]{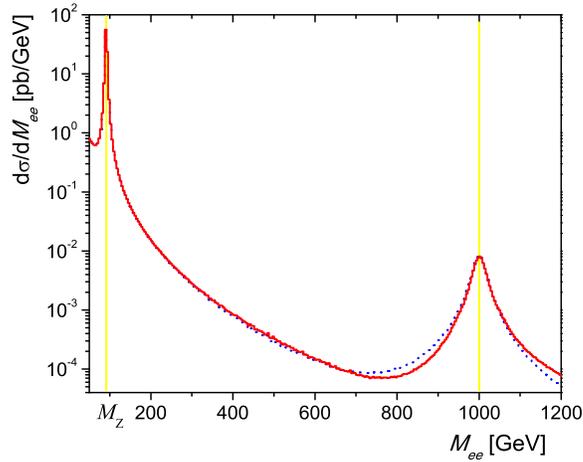}
\caption{\label{fig:1} The invariant dilepton mass distributions for
the $Z'$ boson (dotted) and the chiral excited $Z^*$ boson (solid)
with the Drell--Yan SM background (from the photon and the $Z$
boson) at the LHC for $\sqrt{s}=10$~TeV.}
\end{figure}

Since the current
direct constraints
from the D0 and CDF collaborations
place a lower bound
%
on the mass of
new heavy neutral resonances decaying into light lepton pairs about
1 TeV, we set $M \ge 1$~TeV. For the high
dilepton masses the cross sections of the new boson productions with
$M=1$~TeV at the peaks are about two order of magnitude higher than
the corresponding Drell--Yan background (Fig.~\ref{fig:1}).

\noindent Therefore, the peak should be clearly visible.
For an estimation of the statistical significance of expected signal
we can use the simplest ``number counting'' approach, which is based
on the expected rate of events for the signal, $s$, and background
processes, $b$. The significance can be calculated by the formula
\begin{equation}\label{significance}
    S_{cL}=\sqrt{2\left((s+b)\ln\left(1+\frac{s}{b}\right)
    -s\right)}
\end{equation}
according to the method presented in Appendix A of Ref.~\cite{CMS},
which follows directly from the Poisson distribution.

\begin{figure}[htb]\center
\includegraphics[width=0.43\textwidth]{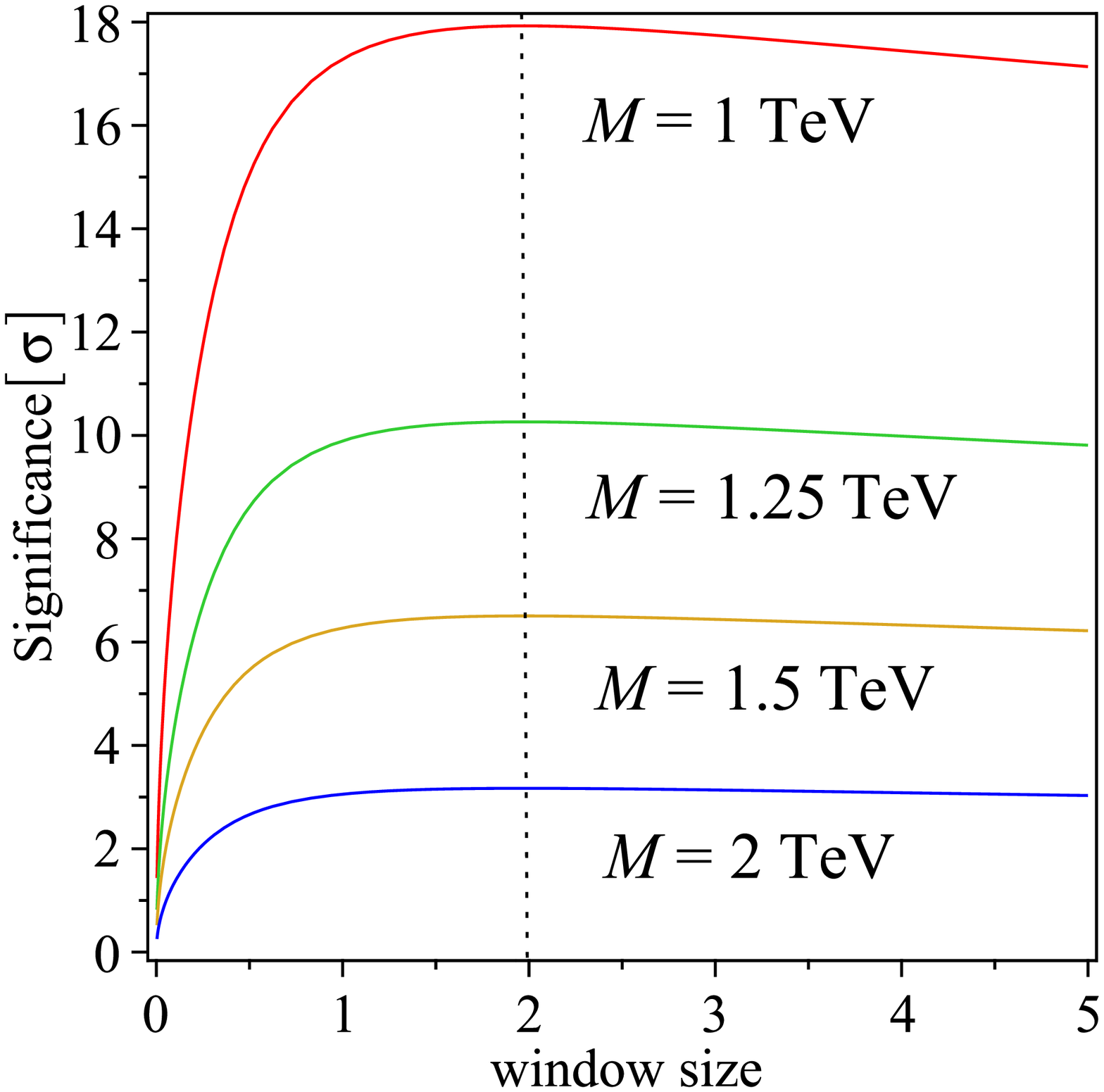}\hspace{0.8cm}
\includegraphics[width=0.43\textwidth]{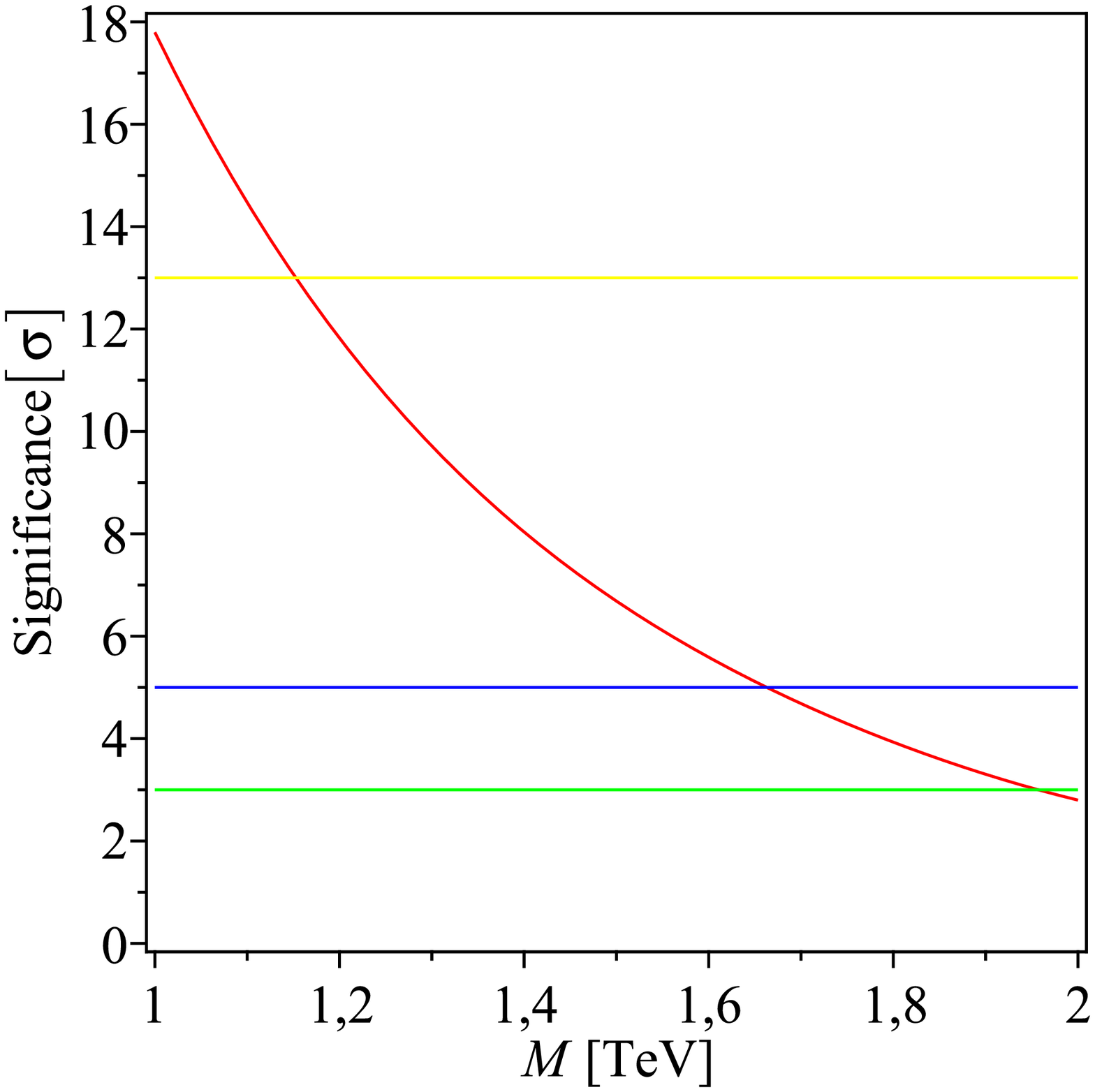}
\caption{\label{fig:significance} Left: The signal significance as
function of
the window size is given for the different masses. Right: The $Z^*$
boson discovery potential at $\sqrt{s}=10$~TeV for 100 pb$^{-1}$.}
\end{figure}

We will focus on the LHC reach with an integrated luminosity of up
to 100~pb$^{-1}$ of data at $\sqrt{s}=10$~TeV. As far as the
center-of-mass energy for the 2010--2011 runs will be 7~TeV, at
which the cross-sections are roughly two times lower, 200~pb$^{-1}$
of data will be equivalent to the first scenario. In order to
estimate the discovery potential and the exclusion limit for the
first LHC data, we need to generate several samples for different
resonance masses. In the ``number counting'' approach, we simply
count the expected number of events within some window under the
resonance including the background. The optimal window size
$[M-2\Gamma,M+2\Gamma]$ can be guessed from the left panel of
Fig.~\ref{fig:significance}.

For different $Z^*$ masses the statistical significance of the expected signal
can be evaluated
using window size $\pm2\Gamma$ around the resonance positions
(the right panel of Fig.~\ref{fig:significance}).
The lowest horizontal line in this plot corresponds to 3$\,\sigma$
level and shows the evidence for discovery, which can be obtained
for the resonance masses up to 2~TeV. The middle horizontal line
shows the discovery potential at 5$\,\sigma$ level for the masses of
the chiral bosons up to 1.65~TeV.

\begin{figure}[htb]\center
\includegraphics[width=0.48\textwidth]{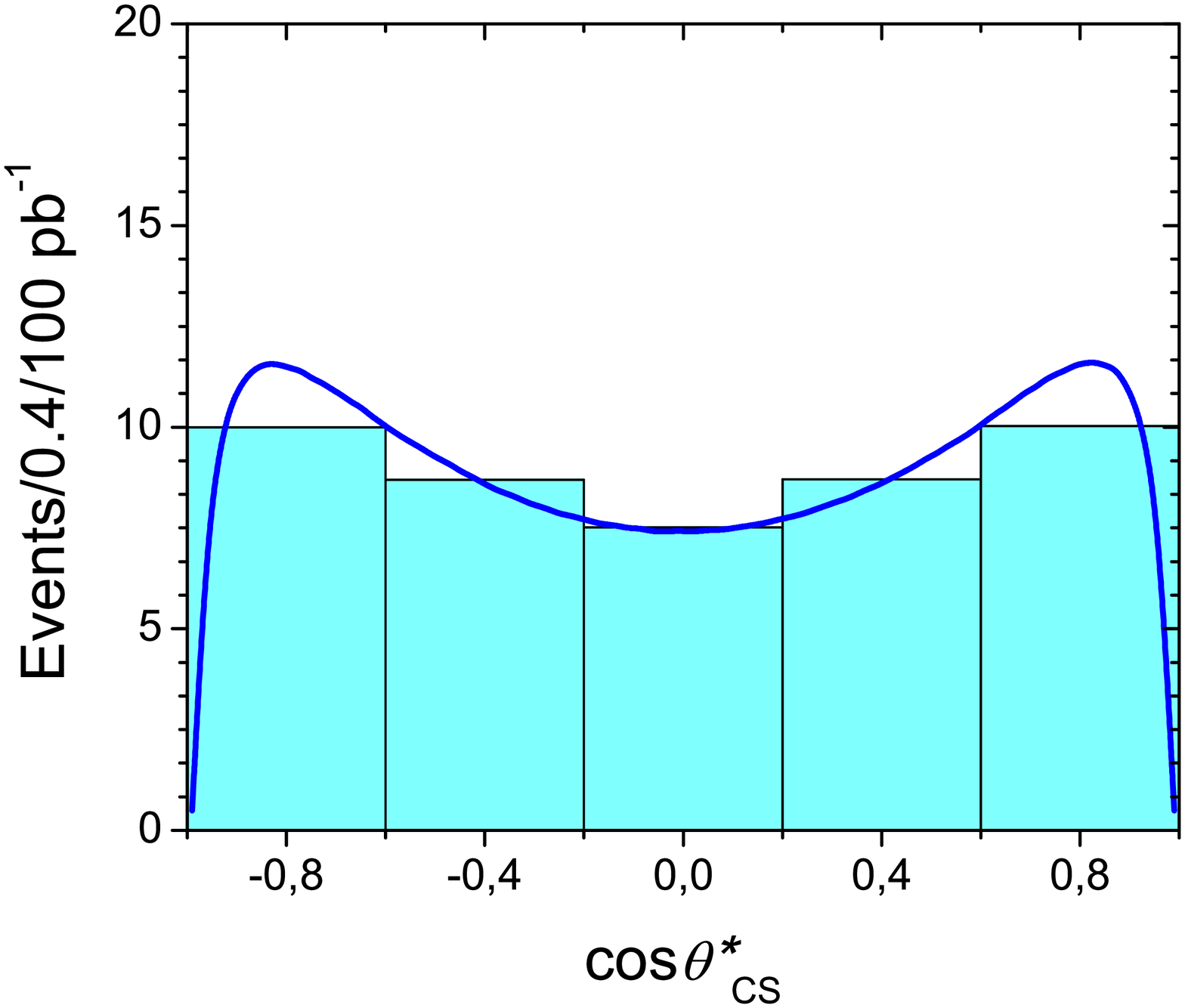}
\includegraphics[width=0.48\textwidth]{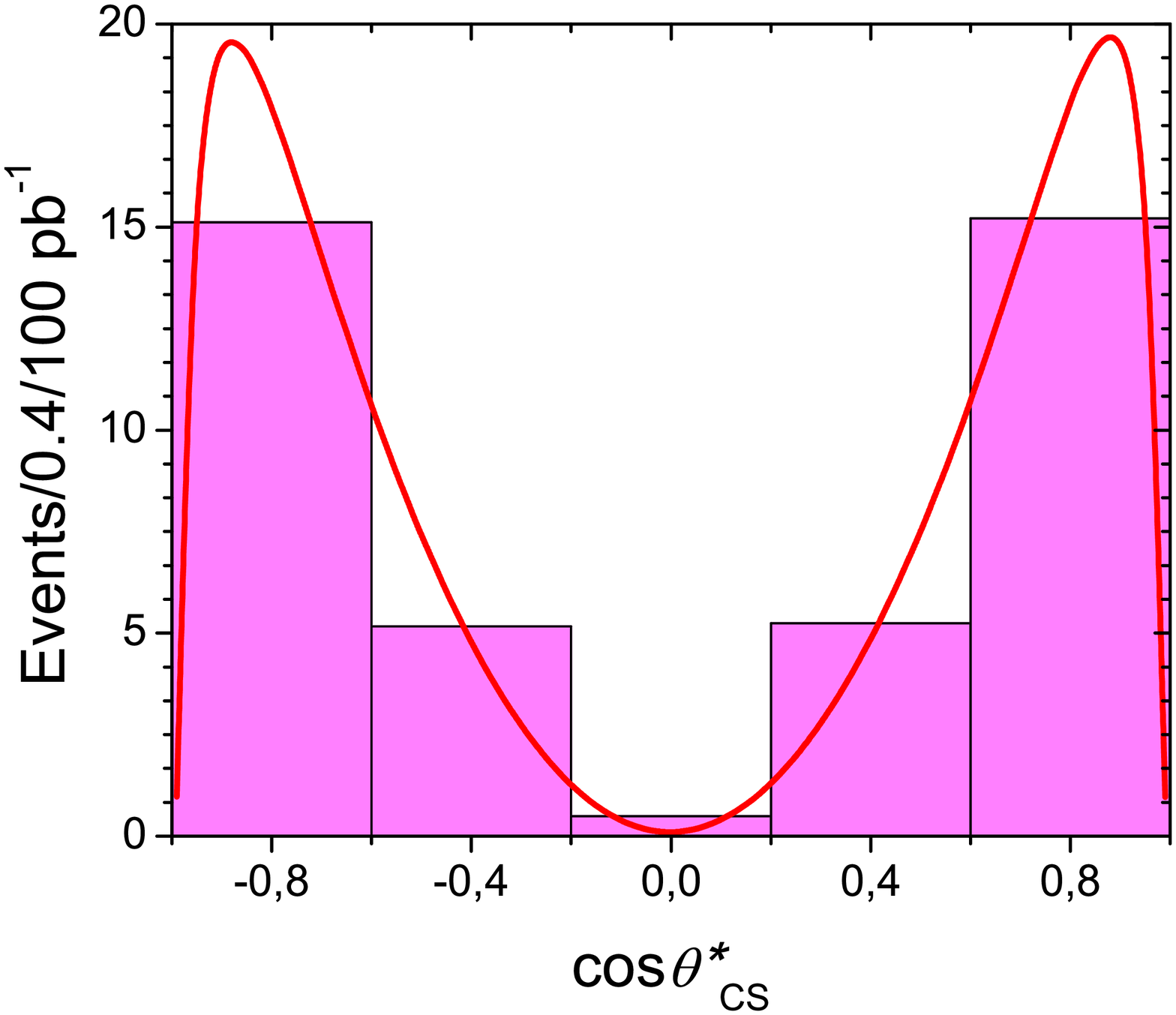}
\caption{\label{fig:CS} The differential distributions of the gauge
$Z'$ boson (left) and the chiral excited $Z^*$ boson (right) as
functions of $\cos\theta^*_{\rm CS}$ for $M=1$~TeV.}
\end{figure}

 The peaks in the invariant mass distributions
originate from the Breit–-Wigner propagator form, which is the same
both for $Z'$ and $Z^*$ bosons in the leading Born approximation.
Therefore, in order to discriminate them we need to investigate
additional distributions selecting only ``on-peak'' events with the
invariant dilepton masses in the chosen range
$[M-2\Gamma,M+2\Gamma]$.
According to 
paper~\cite{brasil} a crucial difference between the chiral bosons
and other resonances should come from the analysis of the angular
distribution of the final-state leptons with respect to the boost
direction of the heavy boson in the rest frame of the latter (the
Collins--Soper frame~\cite{CS}) (Fig.~\ref{fig:CS}).

\begin{figure}[htb]\center
\includegraphics[width=0.48\textwidth]{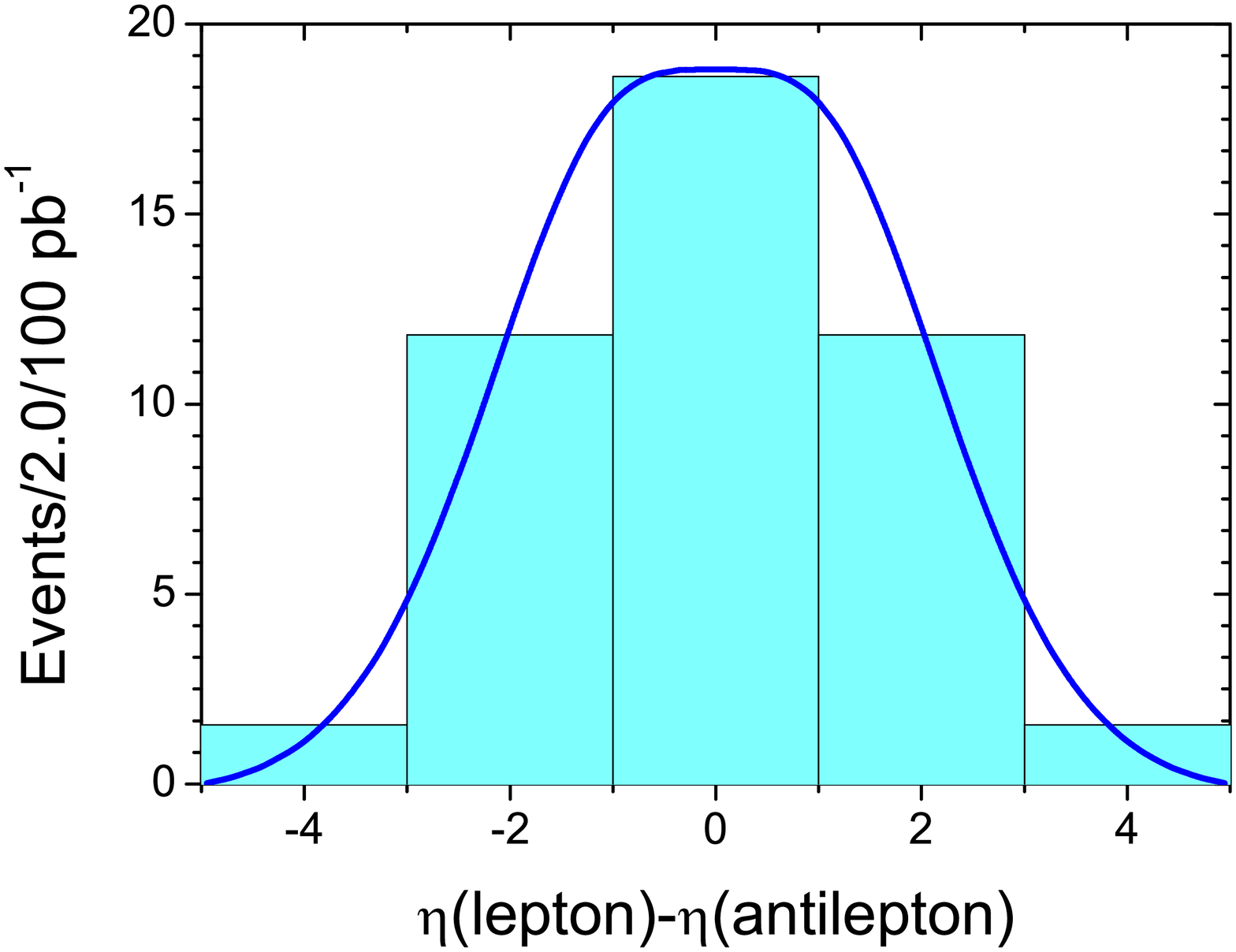}
\includegraphics[width=0.48\textwidth]{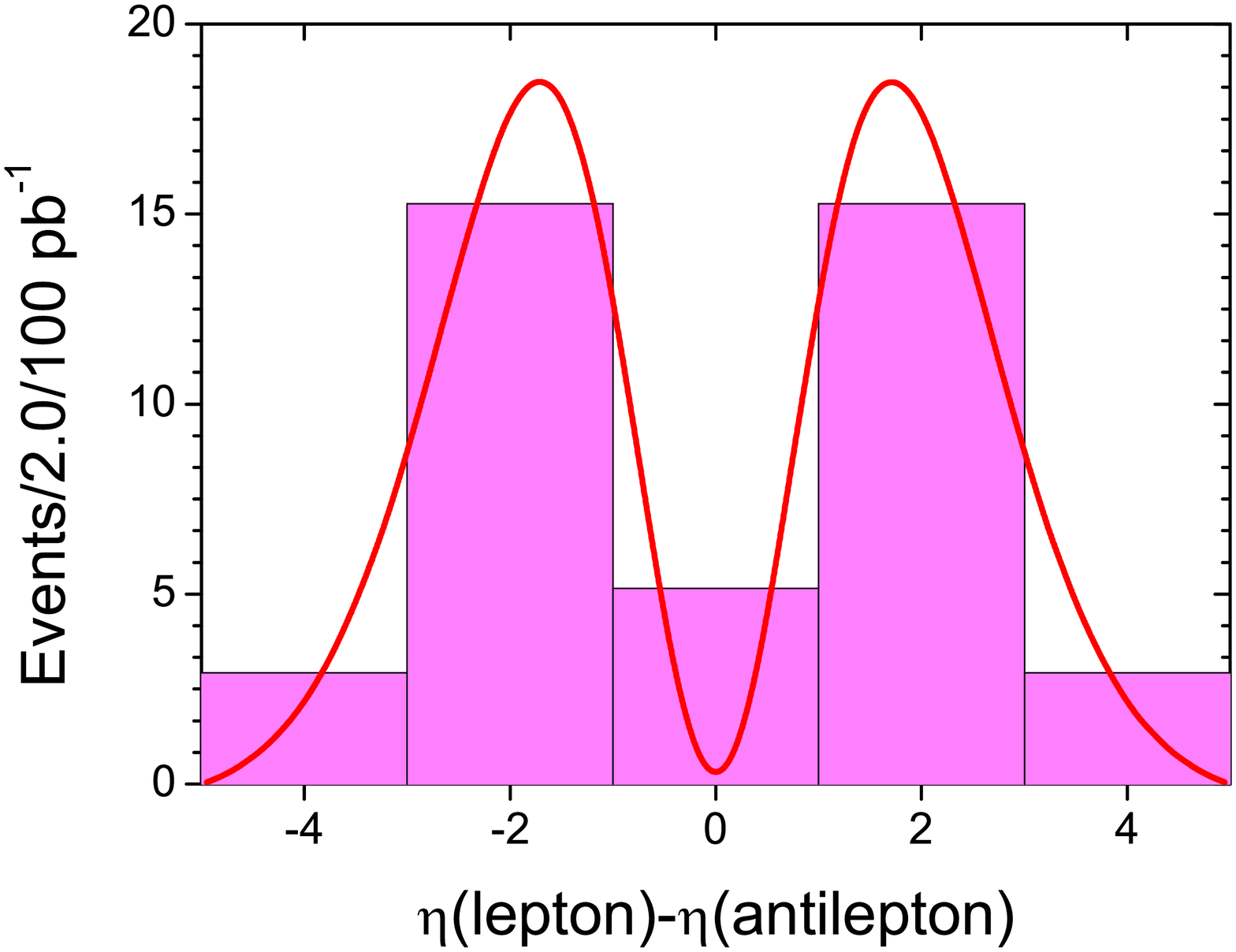}
\caption{\label{fig:eta} The differential distributions for the
gauge $Z'$ boson (left) and the chiral excited $Z^*$ boson (right)
as functions of the difference of the lepton pseudorapidities for
$M=1$~TeV.}
\end{figure}

\noindent Indeed, the angular distribution for the $Z^*$ bosons will
lead to the large negative value of the centre-edge asymmetry
$A_{CE}$~\cite{ACE},
\begin{equation}\label{ACE}
    \sigma A_{CE}=\hspace{-0.2cm}\int^{+\frac{1}{2}}_{-\frac{1}{2}}
    \hspace{-0.2cm}\frac{\drm\sigma}{\drm\cos\theta^*_\sy{CS}}\,\drm\cos\theta^*_\sy{CS}-
    \hspace{-0.1cm}\left[\int^{+1}_{+\frac{1}{2}}
    \hspace{-0.2cm}\frac{\drm\sigma}{\drm\cos\theta^*_\sy{CS}}\,\drm\cos\theta^*_\sy{CS}+
    \hspace{-0.2cm}\int^{-\frac{1}{2}}_{-1}
    \hspace{-0.3cm}\frac{\drm\sigma}{\drm\cos\theta^*_\sy{CS}}\,\drm\cos\theta^*_\sy{CS}
    \right]\!,
\end{equation}
while the distributions of other known resonances (even with
different spins) possess positive or near to zero asymmetries. The
corresponding calculations show that for resonance masses up to
1.15~TeV it is possible to disentangle between the most interesting
cases of $Z^*$ and $Z'$ resonances (horizontal upper line in the
right-hand plot of Fig.~\ref{fig:significance}). 
Another ``unexpected'' consequence of the new angular distribution
is shown in Fig.~\ref{fig:eta}.
Combining these distributions should allow
to differentiate these bosons for higher resonance masses.

To estimate the exclusion limit for given statistics we will apply
simple considerations. For example, if looking for an excess in the
invariant dilepton mass distribution with chosen window above 1 TeV,
we do not find any event (which is in agreement with the SM), then
it is still allowed for 3 signal events to fluctuate down to 0 with
a probability of 5\%. It means that the resonances up to masses
1.65~TeV, which should give more that 3 events, will be excluded at
95\% confidence level.

\section{Remarks on the charged bosons case}

The cleanest method for discovery of the charged heavy bosons at the
hadron colliders is the detection of their subsequent leptonic
decays into isolated high transverse-momentum leptons without a
prominent associated jet activity. In this case they can be observed
through the Jacobian peak in the transverse momentum distribution.
It has become proverbial (see, for example, the
textbook~\cite{Barger}), that the Jacobian peak is characteristic of
all two-body decays. However, it is not the case for the decay of
the new chiral bosons~\cite{trieste}.

It has been found in \cite{two} that tensor interactions lead
to a new angular distribution of the outgoing fermions
\begin{equation}\label{GLR}
    \frac{{\rm d} \sigma(q\bar{q}\to Z^*\!/W^*\to f\bar{f})}
    {{\rm d} \cos\theta} \propto
    \cos^2\theta,
\end{equation}
in comparison with the well-known vector interaction result
\begin{equation}\label{GLL}
    \frac{{\rm d} \sigma(q\bar{q}\to Z'\!/W'\to f\bar{f})}
    {{\rm d} \cos\theta} \propto
    1+\cos^2\theta \, .
\end{equation}
It was realized later~\cite{trieste} that this property ensures a
distinctive signature for the detection of the new interactions at
the hadron colliders. At first sight, the small difference between
the distributions (\ref{GLR}) and (\ref{GLL}) seems unimportant.
However, the absence of the constant term in the first case results
in new experimental signatures.

\begin{figure}[htb]\center
\includegraphics[width=0.48\textwidth]{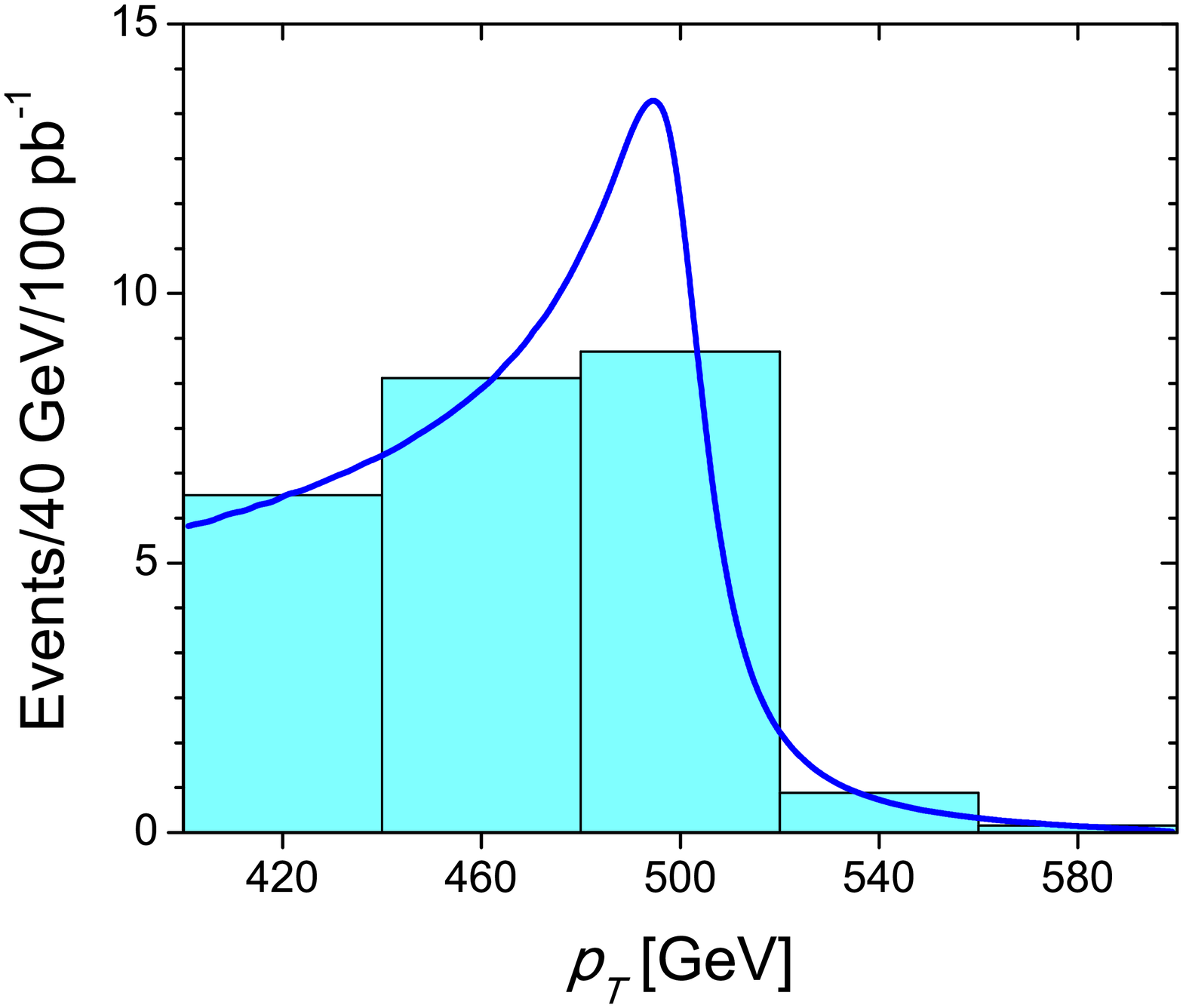}
\includegraphics[width=0.48\textwidth]{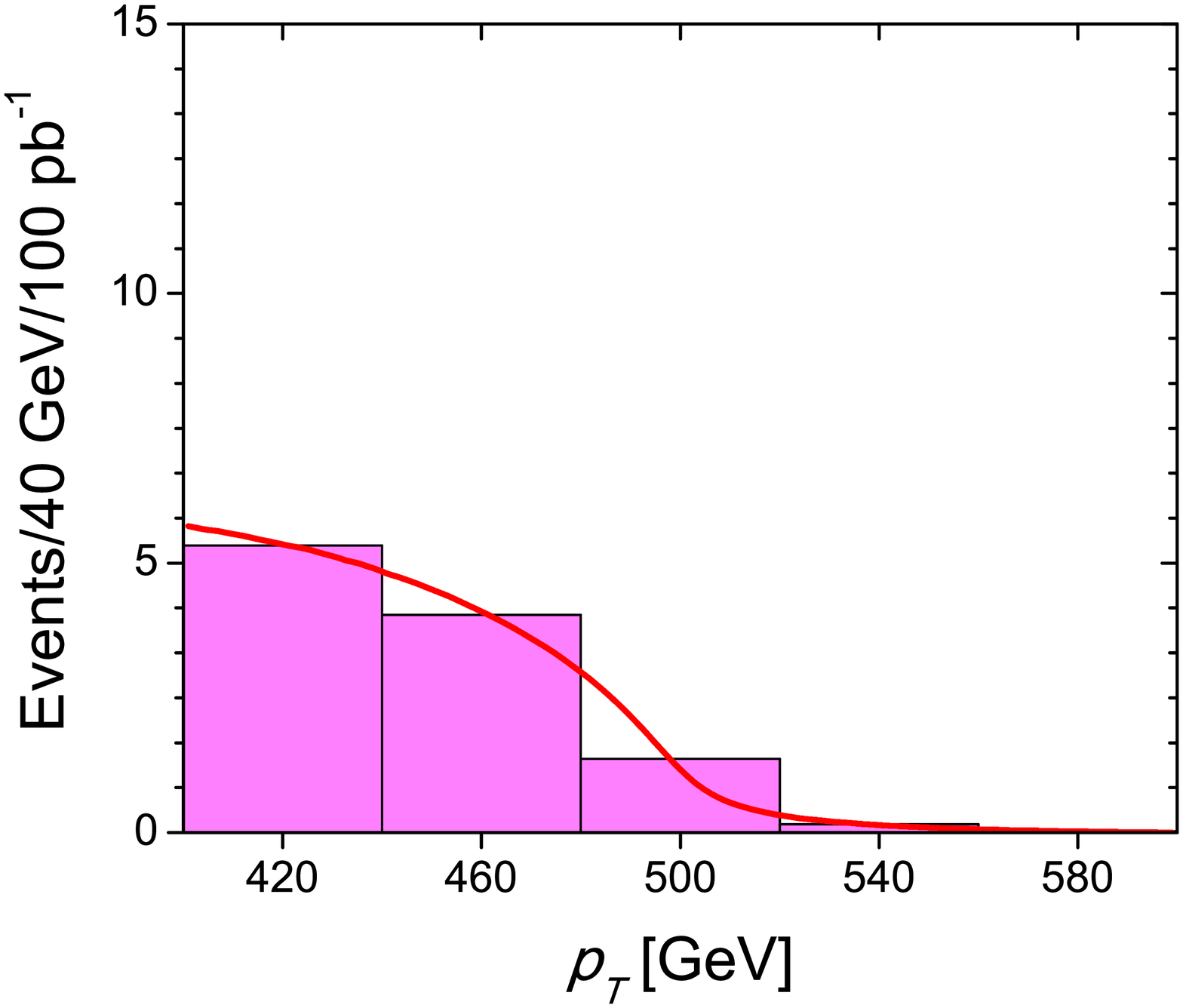}
\caption{\label{fig:pT} The differential distributions for the $Z'$
boson (left) and the chiral excited $Z^*$ boson (right) as functions
of the lepton transverse momentum $p_\sy{T}$ for $M=1$~TeV.}
\end{figure}

The angular distribution for vector interactions (\ref{GLL})
includes a nonzero constant term, which leads to the kinematical
singularity in $p_{\rm T}$ distribution of the final fermion
\begin{equation}\label{1/cos}
    \frac{1}{\cos\theta}\propto\frac{1}{\sqrt{(M/2)^2-p^2_T}}
\end{equation}
in the narrow width approximation $\Gamma <\!\!\!< M$
\begin{equation}\label{narrow}
    \frac{1}{(s-M^2)^2+M^2\Gamma^2}\approx\frac{\pi}{M\Gamma}\delta(s-M^2).
\end{equation}
This singularity is transformed into a well known Jacobian peak due
to a finite width of the resonance. In contrast to this, the pole in
the decay distribution of the $Z^*/W^*$ bosons is canceled out and
the fermion $p_\sy{T}$ distribution even reaches zero at the
kinematical endpoint $p_\sy{T}=M/2$ (Fig.~\ref{fig:pT}).

\noindent The $Z^*/W^*$ boson decay distribution has a broad smooth
hump with the maximum below the kinematical endpoint, instead of a
sharp Jacobian peak. Therefore, in contrast to the usual procedure
of the direct and precise determination of the resonance mass the
new distribution does not allow to do it. Moreover, a relatively
small decay width of the chiral bosons leads to a wide distribution,
that obscures their identification as resonances at hadron
colliders.

\section{Conclusions}

In conclusion we would like to stress
that the new type of spin-1 chiral bosons can exist. They are well
motivated from the hierarchy problem point of view and are predicted
by at least three different classes of theories that represent
different approaches for explaining the relative lightness of the
Higgs doublets. The decay distributions of the chiral bosons differ
drastically from the distributions of the known gauge bosons.
Therefore, the discovery of such type of distributions will point
out to an existence of a compositeness, of a new symmetry and, even,
of extra dimensions.

\acknowledgments The authors thank the organizers, Giorgio
Bellettini, Giorgio Chiarelli, Mario Greco and Gino Isidori for the
invitation and highly appreciate the possibility to present the
results of their work at Les Rencontres de Physique de la Vall\'ee
d'Aoste.

\end{document}